\documentclass[trackchanges,twocolumn]{aastex701}

\usepackage{amsmath}
\usepackage{graphicx}
\usepackage{enumerate}

\begin{document}

\title{A Face-on Accretion Disk Geometry Revealed by Millimeter-wave Periodicity in Sgr A$^*$}

\author[orcid=0009-0006-5208-3962,gname='Kazuki',sname='Yanagisawa']{Kazuki Yanagisawa}
\affiliation{School of Fundamental Science and Technology, Graduate School of Science and Technology, Keio University, 3-14-1 Hiyoshi, Kohoku-ku, Yokohama, Kanagawa 223-8522, Japan}
\email{yanalgm0626@keio.jp}

\author[orcid=0000-0002-5566-0634,gname='Tomoharu',sname='Oka']{Tomoharu Oka}
\affiliation{School of Fundamental Science and Technology, Graduate School of Science and Technology, Keio University, 3-14-1 Hiyoshi, Kohoku-ku, Yokohama, Kanagawa 223-8522, Japan}
\affiliation{Department of Physics, Institute of Science and Technology, Keio University, 3-14-1 Hiyoshi, Kohoku-ku, Yokohama, Kanagawa 223-8522, Japan}
\email{tomo@phys.keio.ac.jp}
 
\author{Ryo Ariyama}
\affiliation{School of Fundamental Science and Technology, Graduate School of Science and Technology, Keio University, 3-14-1 Hiyoshi, Kohoku-ku, Yokohama, Kanagawa 223-8522, Japan}
\email{sirius@keio.jp}

\author{Kazuki Yanagihara}
\affiliation{School of Fundamental Science and Technology, Graduate School of Science and Technology, Keio University, 3-14-1 Hiyoshi, Kohoku-ku, Yokohama, Kanagawa 223-8522, Japan}
\email{lakalakalove.uver@keio.jp}

\author[orcid=0000-0002-9255-4742,gname='Yuhei',sname='Iwata']{Yuhei Iwata}
\affiliation{Mizusawa VLBI Observatory, National Astronomical Observatory of Japan, 2-12 Hoshigaoka, Mizusawa, Oshu, Iwate 023-0861, Japan}
\affiliation{Astronomical Science Program, Graduate Institute for Advanced Studies, SOKENDAI, 2-21-1 Osawa, Mitaka, Tokyo, 181-8588, Japan}
\email{yuhei.iwata@nao.ac.jp}

\author{Mikiya M. Takahashi}
\affiliation{National Institute of Technology, Tokyo College, Hachioji 193-0997, Japan}
\email{m_takahashi@tokyo-ct.ac.jp}
\begin{abstract}

We analyzed 77 epochs of Atacama Large Millimeter/submillimeter Array (ALMA) archival data to investigate flux variability in Sagittarius A$^*$ (Sgr A$^*$), the supermassive black hole at the Galactic Center. Among these, we identified a rare but unusually clear and coherent $\sim$52-minute sinusoidal modulation at 230 GHz, with a statistical significance exceeding 5$\sigma$. Modeling with a Doppler-boosted hotspot scenario yields an orbital radius of $\sim$4 Schwarzschild radii and a disk inclination of 8$^\circ$ (or 172$^\circ$), providing the first direct millimeter wavelength constraint on the inner accretion flow geometry. This nearly face-on inclination is in good agreement with previous constraints from GRAVITY and EHT observations. These findings provide robust, independent evidence that millimeter-wave periodicity can directly probe the innermost accretion flow geometry, offering a powerful complement to variability studies at infrared and X-ray wavelengths.

\end{abstract}


\section{Introduction} 
Sagittarius A$^*$ (Sgr A$^*$), the compact radio source at the Galactic Center, is widely accepted to host a $\sim\!4\!\times\!10^{6}\,M_{\odot}$ supermassive black hole (SMBH) \citep[e.g.,][]{abuter2018detection}.  Sgr A$^*$ exhibits strong variability across a wide range of wavelengths \citep[e.g.,][]{genzel_near-infrared_2003, baganoff2003chandra,subroweit2017submillimeter,michail2024multiwavelength,von2023general,von2025first}. Sgr A$^*$ provides an excellent laboratory for testing general relativity and studying accretion physics at event-horizon scales in the low-luminosity regime. The geometry of Sgr A$^*$ has been constrained through various observational approaches. Comparison of direct imaging of Sgr A$^*$ by the EHT with GRMHD simulations favors a strongly magnetized accretion disk in the magnetically arrested disk (MAD) state, while strongly disfavoring orbital inclinations of $i > 50^{\circ}$ \citep{akiyama2022first}. Polarimetric analysis with ALMA revealed that the observed $\mathcal{Q}–\mathcal{U}$ loop structures are most naturally reproduced by a magnetically dominated accretion disk with a low inclination ($i\sim20^\circ$) and a predominantly vertical magnetic field. GRAVITY observations from 2018 to 2022 further uncovered clockwise loop motions of bright NIR flares with a characteristic period of about 60 minutes at an angular scale of $\sim$100 µas, consistent with a hotspot orbiting at the innermost region of the accretion disk, with an orbital inclination reported to be $i\sim160^\circ$ \citep{abuter2023polarimetry}. Despite this progress, the temporal behavior of Sgr A$^*$ remains poorly understood, particularly the origin of its short-timescale variability.\\\indent
Transient quasi-periodic variabilities in Sgr A$^*$ have been reported across X-ray \citep[e.g.,][]{aschenbach2004x,belanger2006periodic}, infrared \citep[e.g.,][]{genzel_near-infrared_2003,eckart2006flare,michail2024multiwavelength}, and millimeter wavelengths \citep{miyoshi_oscillation_2011}. The characteristic timescales of these variations are on the order of the orbital period near the innermost stable circular orbit (ISCO; $\sim$30 minutes). It should be noted that these variations are distinct from the quasi-periodic oscillations (QPOs) commonly reported in stellar-mass black holes. \citet[I20+ hereafter]{iwata2020time} reported evidence for transient quasi-periodic variability with amplitudes of $\sim\!10\%$ and periods of $\sim\!30$ minutes from a careful analysis of ALMA 230 GHz data. This period is comparable to the orbital period at the innermost stable circular orbit (ISCO) for a non-spinning $\sim\!4\!\times\!10^{6}\,M_{\odot}$ black hole. The reported periodic variations are not persistent but are often transient. Although the origin of such sporadic periodicity remains elusive, it may provide important clues to the geometry, kinematics, and magnetic field structure of the innermost accretion flow.\\\indent
\begin{figure*}[htbp]
\begin{center}
\includegraphics[width=0.7\textwidth]{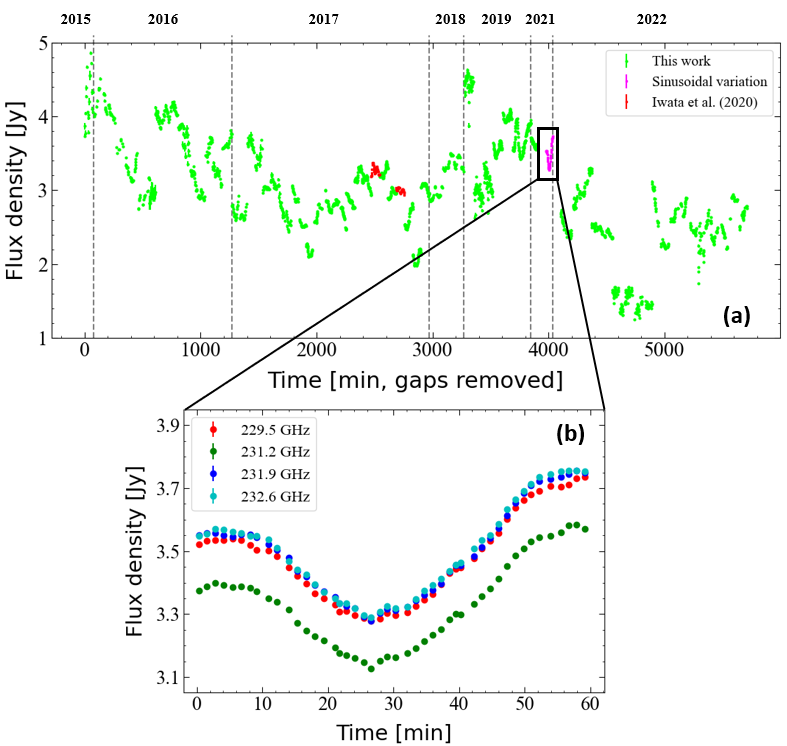}
\caption{(a) Light curves of Sgr A$^*$ obtained over 77 epochs from 2015 to 2022.  The red curves indicate data sets in which periodic variations were identified by I20+, while the magenta curve highlights the periodic variation newly identified in this study. The vertical gray dashed lines indicate the boundaries of the calendar years (UT) when the data were observed.}
(b) Light curve of Sgr A$^*$ observed on 2021 July 22 with ALMA at 229.5 GHz (red), 231.2 GHz (green), 231.9 GHz (blue), and 232.6 GHz (cyan). The horizontal axis shows time in minutes relative to the observation start. 
\label{light}
\end{center}
\end{figure*}
Here, we analyze 230 GHz ALMA observations of Sgr A$^*$ that were selected for long continuous coverage and high cadence, yielding a set of 77 light curves with durations of $\sim$30–70 minutes.
While most of the light curves show no significant periodicity, we report a statistically robust, clear, transient sinusoidal modulation with a period of $\sim$52 minutes in one epoch.
By interpreting this modulation as emission from a compact hotspot orbiting at the inner edge of the accretion disk, we constrained the orbital velocity of the hotspot and the inclination of the disk.
These parameters are consistent with those inferred from NIR flares observed with GRAVITY, supporting a unified picture of hotspot-driven variability across the millimeter, infrared, and X-ray bands.

\section{Observations and Data Reduction} \label{observation}
We systematically surveyed 230 GHz data of Sgr A$^*$ in the ALMA Science Archive.  The selection criteria required (1) focusing on Sgr A$^*$'s field of view, (2) 230 GHz frequency band, (3) angular resolution below $1\arcsec$, and (4) integration time exceeding 1 hour.  Applying these filters, we identified data from 12 projects executed between July 2015 and May 2022 that met our requirements. We applied the self-calibration and CLEAN algorithm to these data using the Common Astronomy Software Applications (CASA) package \citep[e.g.,][]{mcmullin2007casa,bean2022casa}.\\\indent
The raw ALMA visibility data were calibrated using the standard scriptForPI.py provided by the observatory, which performs flagging and applies bandpass, flux, and phase calibrations using the appropriate calibrators. After this standard calibration, we performed iterative phase self-calibration on Sgr A*, following the method described in I20+. For each iteration, we executed the \texttt{tclean} task to create an initial model image of Sgr A*. We then applied iterative phase self-calibration using \texttt{gaincal} and \texttt{applycal}. We started with a solution interval of \texttt{inf} (per scan), and subsequently shortened it to int (per integration), improving the coherence of the data. This procedure improved the dynamic range of the images and stabilized the measured flux density variations. After performing self-calibration, we applied the CLEAN algorithm to the data to deconvolve the synthesized beam pattern and produced noise-reduced snapshot images. The flux density and its associated uncertainty were obtained by performing fits to the snapshot images with the CASA task \texttt{imfit}.\\\indent
The ALMA Cycle 7 observations (Project 2019.1.01559.S; PI: L. Murchikova) discussed in this paper were carried out on 2021 July 20 and 22. The four spectral windows (SpWs) were centered at 229.5, 231.2, 231.9, and 232.9 GHz, each with a bandwidth of 2 GHz. Owing to the presence of strong and complex absorption feature, which could correspond to CO $J$=2--1 line, the 231.2 GHz data were excluded from further analysis.


\section{Variability analysis} \label{results}
We obtained 77 light curves, each with a duration of $30\mbox{--}70$ minutes and the data are binned to $\sim$1 min, covering a total of 90 hours (Figure \ref{light}a). Note that observational gaps between epochs are removed.

\subsection{Period Analysis}
\begin{figure}[htbp]
\begin{center}
\includegraphics[width=0.5\textwidth]{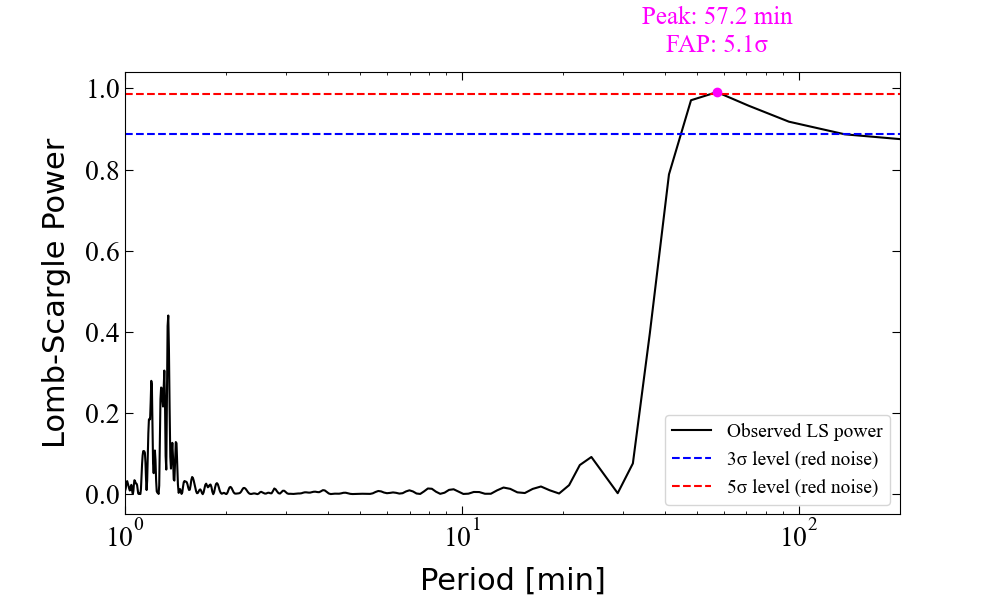}
\caption{Generalized Lomb–Scargle periodogram of the light curve observed on 2021 July 22.} The black line shows the LS power of the observed light curve in this study. The false-alarm probability (FAP) was estimated by generating $10^5$ red-noise light curves and is indicated at the 3$\sigma$ (blue dashed line) and 5$\sigma$ (red dashed line) levels. Peaks in the LS power are marked with magenta points, with the corresponding FAP values indicated simultaneously.
\label{GLS}
\end{center}
\end{figure}

To search for periodic signals, we applied the Generalized Lomb-Scargle (GLS) periodogram \citep{zechmeister2009generalised} to each light curve after subtracting a linear trend to remove slow variations. For a set of observational data points $(t_i, y_i)$, the GLS is defined as:

\begin{multline*}
p(\omega)=\frac{1}{YY\cdot D}\big[SS\cdot YC^2+CC\cdot YS^2-2CS\cdot YC\cdot YS\big]
\label{eq:GLS_def}
\end{multline*}
with:
\begin{equation*}
D(\omega)=CC\cdot SS-CS^2
\label{eq:GLS_D}
\end{equation*}
where $w_{\rm i}$ is the probability weight, and $Y, S$ and $C$ denote the respective sums.

\begin{equation*}
\begin{split}
w_{\rm i}&=\frac{1}{W}\frac{1}{\sigma_{\rm i}^2}\:\:\:\:\:\:\Big(  W=\sum\frac{1}{\sigma_{\rm i}^2} \Big)\\
Y&=\sum w_{\rm i}y_{\rm i}\\
C&=\sum w_{\rm i}\cos{\omega t_{\rm i}}\\
S&=\sum w_{\rm i}\sin{\omega t_{\rm i}}\\
\end{split}
\label{eq:GLS_W}
\end{equation*}
The false-alarm probability (FAP) of each periodogram peak was estimated from $10^{5}$ Monte Carlo realizations of red-noise light curves with a spectral density index of $\alpha=2.20$ ($\mathrm{PSD}\propto f^{-2.20}$; \citealt{weldon2023near}). These red-noise light curves were constructed with the same temporal sampling as our observed data.  We considered a signal significant if $\mathrm{FAP}\!<\!2.9\!\times\!10^{-7}$, corresponding to $>\!5\sigma$ significance for Gaussian noise. Periodic variability was detected in three epochs, two of which occurred in 2017 and were already reported by I20+. The periodogram of the 2021 light curve, which also exhibits a significant peak, is shown in Figure \ref{GLS}. The remaining 74 light curves show no comparable periodicity above the threshold, suggesting that such strong, coherent millimeter-wave periodic signals are rare or transient phenomena in Sgr A$^*$.

\subsection{Detection of a Sinusoidal Variation}
The newly detected periodicity was identified in the data obtained on 2021 July 22 during ALMA Cycle 7 observations (project 2019.1.01559.S; PI: L. Murchikova).  It exhibited a significant peak at $P\!\sim\!50$ minutes with a extremely small false alarm probability ($\mathrm{FAP}\!<\!1.8\times10^{-7}$).  The corresponding light curve displays a remarkably clear and unambiguous sinusoidal modulation with a period of approximately 52 minutes (Figure \ref{light}b), which remains coherent over the $\sim\!60$-minute duration of the observation.  This distinct sinusoidal pattern is consistently present across three of the four analyzed spectral windows. During this epoch, Sgr A$^*$ maintained an average activity state, with flux densities ranging from 3.3 to 3.7 Jy. The amplitude of the observed sinusoidal variation is approximately 14\%, consistent with previously reported periodic variations. Taken together, these results strongly suggest that the observed modulation originates from the relativistic beaming of a rapidly orbiting ``hotspot".

\section{Modeling} \label{dis}
\begin{figure}[t]
\begin{center}
\includegraphics[width=0.5\textwidth]{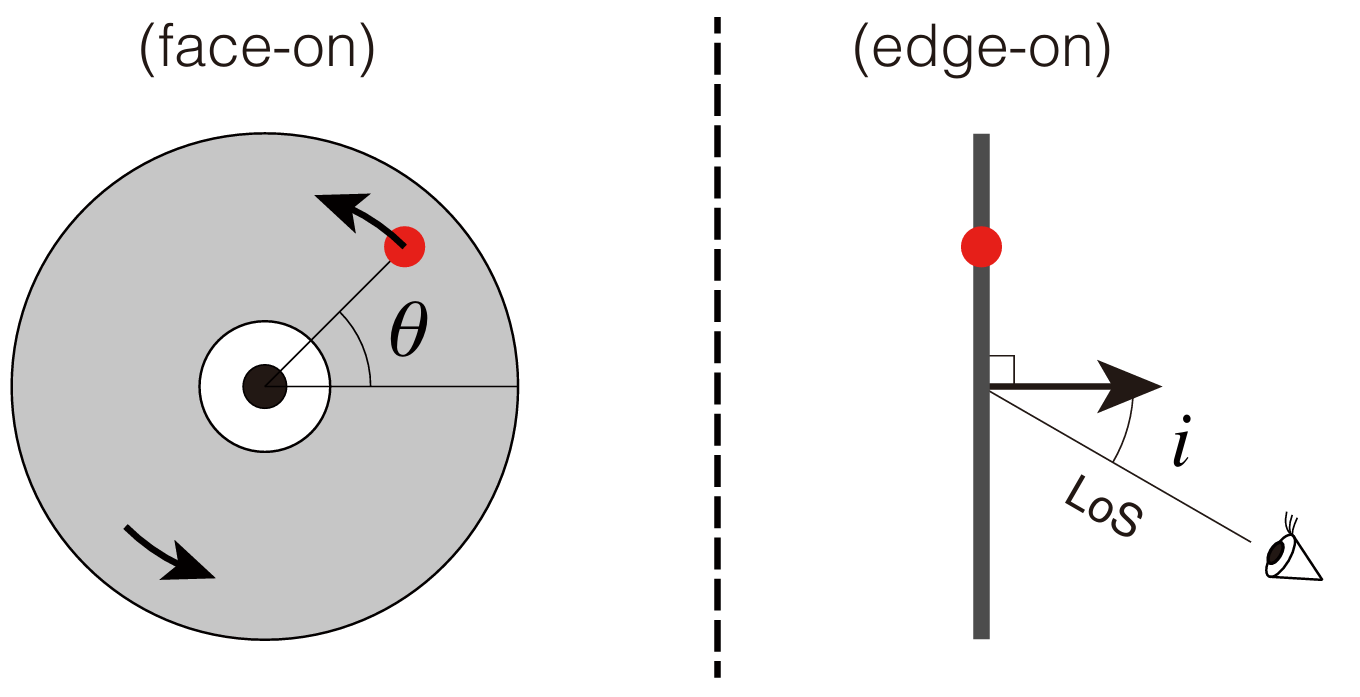}
\caption{Schematic diagram of the hotspot model. The left panel illustrates a face-on view of the accretion disk surrounding Sgr A$^*$, while the right panel shows an edge-on perspective. The angle $\theta$ denotes the azimuthal position of the hotspot along its orbit, and $i$ represents the inclination angle between the disk's angular momentum vector and the line of sight (LoS). }
\label{hotspot}
\end{center}
\end{figure}

\begin{figure}[htbp]
\begin{center}
\includegraphics[width=0.5\textwidth]{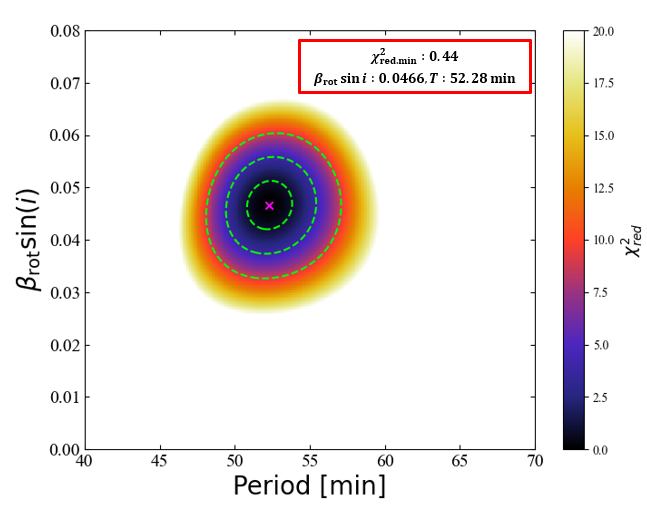}
\caption{The $\chi^2$ contour obtained by fitting the hotspot model to the flux density at 229.5 GHz.  The magenta point marks the minimum of the reduced chi-square, $\chi^2_{\text{red,min}}$ , corresponding to $T\!=\!52.28$ min and $\beta_{\rm rot}\sin{i}\!=\!0.0466$. The lime contour indicates the confidence regions spanning $1\sigma$ to $3\sigma$.}
\label{chi}
\end{center}
\end{figure}

\begin{figure}[htbp]
\begin{center}
\includegraphics[width=0.5\textwidth]{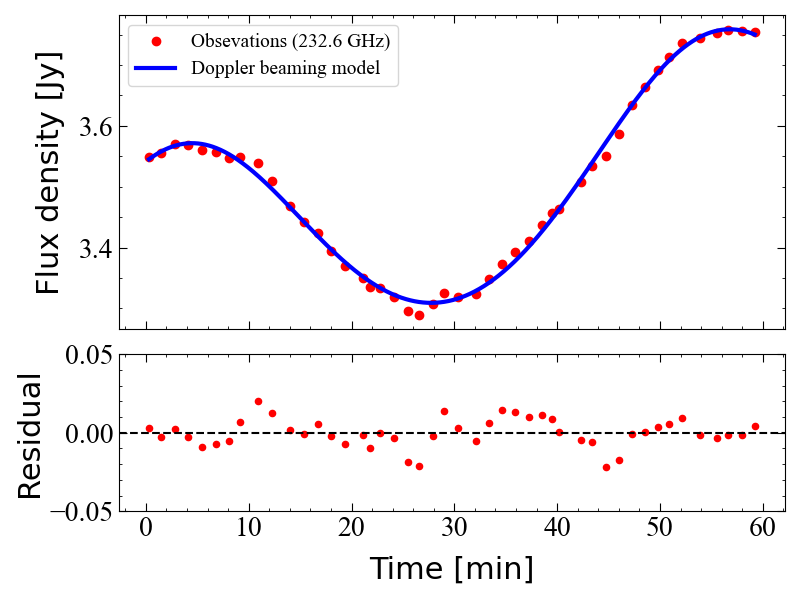}
\caption{Top: Fitting of the Doppler-boost model to observational data.  The best-fit parameters $(T_{\text{min}}, \beta_{\text{min}}\sin{i}_{\text{min}})$, corresponding to the minimum reduced chi-square $\chi^2_{\text{red,min}}$, are shown with the blue curve, while the observed flux density at 232.6 GHz is plotted with red. Bottom: Residuals between the model and observations. With residuals below 1\% of the sinusoidal amplitude, the model shows strong consistency with the observations. The 232.6 GHz light curve is shown as a representative example, as the same sinusoidal modulation is consistently seen in the other spectral windows.}
\label{fit_res}
\end{center}
\end{figure}

To quantify the properties of the orbiting hotspot, we adopted a circular-orbit hotspot plus accretion disk model. Figure \ref{hotspot} presents a schematic diagram of this model. Within the framework, the flux density of Sgr A$^*$ is expressed as:
\begin{equation}
F(t)=\frac{A}{\left\{ 1-\beta_{\rm rot}\sin{i}\cdot\cos\left(2\pi\,t/T-\phi \right) \right\}^3} + B + C\,t
\label{eq:doppler_1}
\end{equation}
where $\beta_{\rm rot}\!\equiv\! v_{\rm rot}/c$ denotes the rotational velocity of the hotspot normalized by the speed of light, $i$ is the orbital inclination, $\phi$ is phase correction term, and $T$ is the orbital period. A least-squares fit of Eq.(\ref{eq:doppler_1}) to the observed sinusoidal light curve yielded well-constrained best-fit values for the parameters $A$, $B$, $C$, $T$, $\phi$, and $\beta_{\rm rot}\sin{i}$. This expression is derived by rearranging equation (4.97b) of \citet{rybicki2024radiative}. Figure \ref{chi} shows the $\chi^2$ contours as a function of $T$ and $\beta_{\rm rot}\sin{i}$, while Figure \ref{fit_res} presents the fitted light curve. Both parameters are tightly constrained by the data, giving $T\! =\! 52.28^{+1.46}_{-1.54}\,\mbox{min}$ and $\beta_{\rm rot}\sin{i}\! =\!(4.7\!\pm\!0.5)\!\times\!10^{-2}$.

Leveraging these precisely determined values and applying Kepler's third law for motion around a spinning black hole with a black hole mass of $4.297\!\times\! 10^6\, M_{\sun}$ \citep{abuter2022mass}, we derived the rotational velocity, orbital radius, and crucial orbital inclination.  In this model, however, the orientation of the black hole's angular momentum vector remains unconstrained, leading to a two-fold degeneracy in the inferred inclination angle.
\begin{equation}
T=2\pi\frac{GM}{c^3}\Big[\Big(\frac{2r}{r_{\text{s}}}\Big)^\frac{3}{2}+a_*\Big] = \frac{2\pi r}{v_{\rm rot}}\, ,
\label{eq: Kep2}
\end{equation}
where $G$ is the gravitational constant, $M$ is the black hole mass, $r$ is the orbital radius, $r_{\text{s}}\!\equiv\!2GM/c^2$ is the Schwarzschild radius, and $a_*$ is the dimensionless spin parameter of the black hole.  This framework allows the physical parameters of the innermost accretion flow to be directly inferred from the observed periodic modulation.

By fitting Eq.(\ref{eq:doppler_1}) to the observed light curve, we obtained the projected velocity of the hotspot, $\beta_{\rm rot}\sin{i}$, together with the orbital period, $T$. Substituting this period into Eq.(\ref{eq: Kep2}) yields the orbital radius $r$ for various values of the dimensionless spin parameter $a_*$, ranging from 0 (non-spinning BH) to 1 (maximally spinning BH). From the relation defined in  Eq.(\ref{eq: Kep2}), the orbital radius $r$ and period $T$ provide the rotational velocity of the hotspot, $v_{\rm rot}$, which in turn gives $\beta_{\rm rot} = v_{\rm rot}/c$.  Finally, by comparing this derived $\beta_{\rm rot}$ with the fitted $\beta_{\rm rot}\sin{i}$, we determine the orbital inclination angle $i$.

\begin{figure*}[htbp]
\begin{center}
\includegraphics[width=0.9\textwidth]{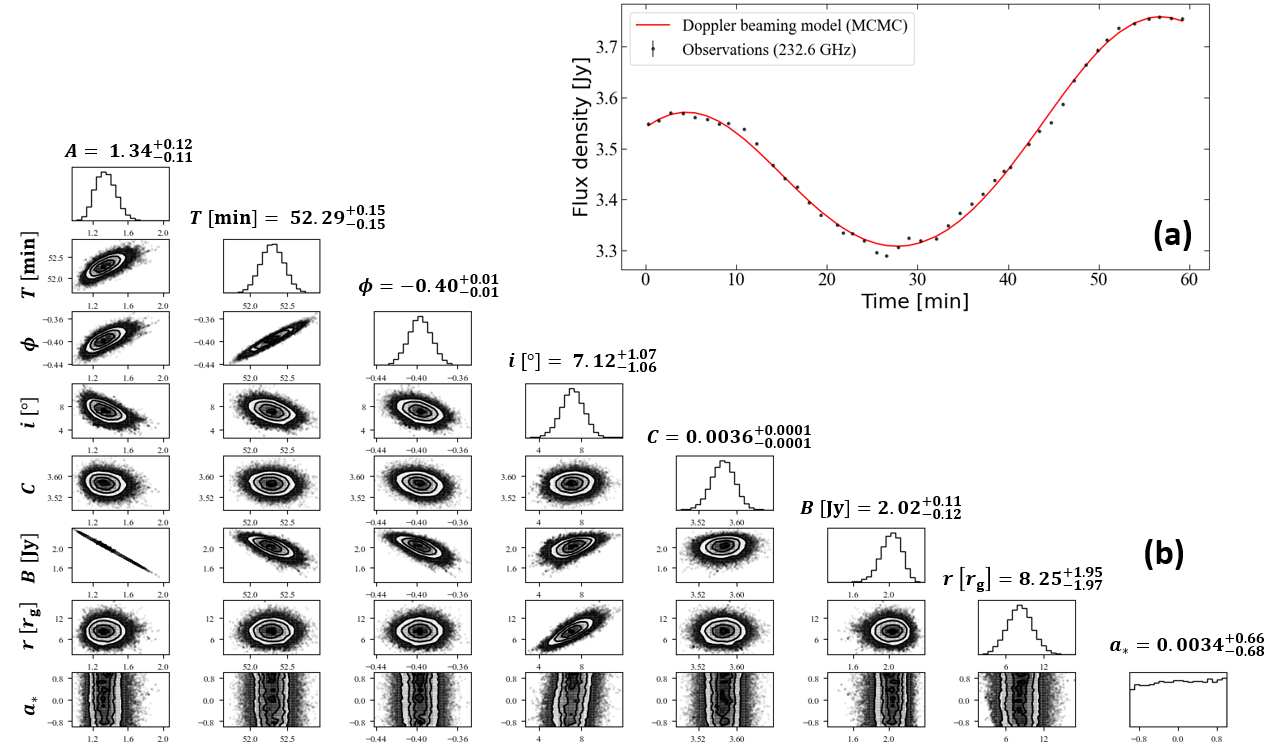}
\caption{(a) Light curve fitted with MCMC. Gray points represent the 232.6 GHz observations, and the red line shows the fit using Eq.(3).
(b) Corner plot for all eight parameters. Each panel displays the correlations between parameters, with the posterior distributions shown along the top. Contours correspond to 1$\sigma$, 2$\sigma$, and 3$\sigma$ levels, and the indicated uncertainties represent 1$\sigma$. The posterior distribution of $a_*$ is flat, indicating that this parameter cannot be constrained by the current model. Note that $r$ is expressed in units of $r_{\rm g}$.}
\label{MCMC}
\end{center}
\end{figure*}

For a non-spinning Schwarzschild black hole ($a_*\!=\!0.0$), our calculations yield $v_{\rm rot}\! =\! (3.48\!\pm\! 0.03)\!\times\! 10^{-1}\, c$ , an orbital radius of $r\! =\! (4.1\!\pm\! 0.1)\, r_{\text{s}}$ (where $r_{\text{s}}$ is the Schwarzschild radius), and an exceptionally shallow inclination angle of $i\! =\! 7\fdg 7\!\pm\! 0\fdg 8$ or $172\fdg 3\!\pm\! 0\fdg 8$.  Assuming a maximally spinning Kerr black hole ($a_*\!=\!+1.0$) gives $v_{\rm rot}\! =\! (3.39\!\pm\! 0.03)\!\times\! 10^{-1}\, c$ , $r\! =\!  (4.0\!\pm\! 0.1)\, r_{\text{s}}$, and $i\! =\! 7\fdg 9\pm 0\fdg 8$ or $172\fdg 1\pm 0\fdg 8$ (Table \ref{table}).  Even for a counter-rotating Kerr black hole ($a_*\!=\!-1.0$), the parameters remain consistent: $v_{\rm rot}\! =\! (3.58\!\pm\! 0.04)\!\times\! 10^{-1}\, c$ , $r\! =\!  (4.2\!\pm\! 0.1)\, r_{\text{s}}$, and $i\! =\! 7\fdg 5\!\pm\! 0\fdg 8$ or $172\fdg 5\!\pm\! 0\fdg 8$.  Notably, these orbital parameters are determined with significantly higher precision compared to those inferred from infrared hotspots \citep{abuter2018detection}.\\\indent
Furthermore, by applying a Markov chain Monte Carlo (MCMC) analysis to the combined form of Eq.(\ref{eq:doppler_1}),(\ref{eq: Kep2}), we simultaneously constrained not only the parameters of Eq.(\ref{eq:doppler_1}) but also $r$, $a_*$:
\begin{multline}
F(t)=\frac{A}{\left\{ 1-\frac{r}{r_{\rm g}}\Big[\big(\frac{r}{r_{\rm g}}\big)^{3/2}+a_*\Big]^{-1}\sin{i}\cdot\cos\left(\frac{2\pi\,t}{T}-\phi \right) \right\}^3} \\
+ B + C\,t
\label{eq:doppler_MCMC}
\end{multline}
where $r_{\rm g}=r_{\rm s}/2=GM/c^2$ is the gravitational radius. The corresponding corner plot is shown in Figure \ref{MCMC}. We adopted wide, uninformative priors for all parameters: $a_*$ and $\phi$ were assigned uniform priors, while the remaining parameters used Gaussian priors centered on the best-fit values from our initial analysis, reflecting prior knowledge without biasing the posterior. While the posterior distribution of the spin parameter remains flat, the other parameters are consistent with the results obtained from the analysis using Eq.(\ref{eq:doppler_1}),(\ref{eq: Kep2}). The flat posterior of the spin can be attributed to the weak dependence on $a_*$ in Eq.(\ref{eq: Kep2}), consistent with the fact that the inferred value of $i$ did not exhibit significant changes across different values of $a_*$. We note that this analysis does not take into account general relativistic (GR) effects or the finite size of the hotspot; thus, the reported parameter uncertainties represent only statistical errors. The impact of GR effects and hotspot size on the light curves will be discussed in Section \ref{dis}.

\begin{table}[htbp]
\raggedright
\caption{Hot spot parameters for each spin}
\vspace{-5pt}
\begin{tabular}{cccc} \hline
     $a_*$ & $r$ [$r_{\rm s}$] & $\beta_{\rm rot}$ [$c$] & $i$ [$^\circ$]\\ \hline
     $1$ & $4.0\pm0.1$ &  $0.339\pm0.003$ &$7.9^{\circ}\pm0.8^{\circ}$ or $172.1^{\circ}\pm0.8^{\circ}$\\ 
      $0$&$4.1\pm0.1$ &  $0.348\pm0.003$ &$7.7^{\circ}\pm0.8^{\circ}$ or $172.3^{\circ}\pm0.8^{\circ}$  \\
     $-1$ & $4.2\pm0.1$ &  $0.358\pm0.004$ &$7.5^{\circ}\pm0.8^{\circ}$ or $172.5^{\circ}\pm0.8^{\circ}$ \\ \hline
   \end{tabular}
  \label{table}
\end{table}

\section{Discussion}\label{dis}
\subsection{Properties of the Periodic Variability}
Although no flares have been reported around the epoch of this sinusoidal modulation in other studies, Figure \ref{light}(a) suggests that Sgr A* was relatively active even during its quiescent phase. Our analysis reveals a highly coherent periodicity of $\sim$52 min. This period is longer than those previously reported (I20+: 30 min, \citet{abuter2018detection}: 45 min, \citet{michail2024multiwavelength}: 35 min), which can be interpreted as arising from differences in the orbital radius of the hotspot. In paticular, our analysis demonstrates that the millimeter-wave hotspot detected with ALMA is robustly orbiting at $\sim\!4$ Schwarzschild radii with a mildly relativistic velocity in an almost face-on accretion disk. Furthermore, transient and coherent periodic variations, such as those reported in this study and by I20+, are very rare. This can be understood under the assumption that, in most epochs, multiple hotspots contribute to the millimeter emission, and epochs in which a single hotspot dominates are rare.  \\\indent 
The orbital parameters are determined with exceptional precision, exceeding those inferred from previous infrared hotspot studies, and are fully consistent with the values suggested by recent studies on the inclination of Sgr A$^*$  \citep{abuter2018detection,issaoun2019size, akiyama2022first}. Notably, the inferred orbital radius lies just outside the innermost stable circular orbit (ISCO) of a non-spinning black hole and is comparable to the emission ring diameter resolved by the EHT. This compelling agreement indicates that the observed millimeter-wave periodicity arises from the innermost accretion flow and shares a common physical origin with the mechanisms producing infrared and X-ray flares.

\subsection{Impact of General Relativistic Effects}
In Eq.(\ref{eq: Kep2}), we adopted a simplified model that neglected both the finite size of the hotspot and general relativistic (GR) effects (e.g., light bending and gravitational redshift). To assess the possible impact of these systematics, we performed general relativistic radiative transfer (GRRT) calculations using the code \texttt{CARTOON} \citep{takahashi20223d}. The orbital inclination and period were fixed to the values obtained in this study, and the SMBH mass was set to $4.297\times10^6\,M_{\odot}$. We simulated light curves for four models, corresponding to two spin values ($a_*\in \{0,0.998\}$) and two hotspot radii ($r\in\{0.5, 1.0\}\,r_{\rm g}$). In all cases, the simulated light curves differed from those predicted by the Doppler-beaming model (Eq.\ref{eq:doppler_1}) by less than 2\% of this amplitude. This indicates that the systematic uncertainties introduced by neglecting GR effects and finite hotspot size are small. Indeed, distortions of light curves due to relativistic effects have been reported to be very small for nearly face-on accretion flows  \citep{von2023general}, consistent with the nearly sinusoidal variations obtained in this study. We modeled the hotspot as a finite-size emitter but did not include possible temporal variations in its size, emission mechanism, or intrinsic physical properties during an orbit. Such time-dependent changes could have a more significant impact on the observed modulation and inferred parameters.
Developing a more comprehensive model that accounts for these effects will be an important subject for future study.\\\indent
Taken together, these results strongly support a unified picture in which transient, relativistically orbiting hotspots near the ISCO play a central role in generating the observed multi-wavelength variability of Sgr A$^*$.  Building on this interpretation, we highlight the broader implications of our findings for future observational and theoretical efforts.

\section{conclusion}
In this work, we analyzed 77 epochs of ALMA archival data to investigate flux variability in Sgr A$^*$. Our main findings are:

\begin{enumerate}
\item We identified a transient and remarkably coherent $\sim$52-minute sinusoidal modulation with $>5\sigma$ statistical significance, representing among the clearest millimeter-wave periodic signals yet observed from Sgr A$^*$.
\item Modeling with a Doppler-boosted hotspot scenario yielded an orbital period of $\sim$52 min and a projected velocity $\beta\sin{i}\sim$0.047, corresponding to an orbital radius of $\sim$4 $r_{\rm s}$ and an almost face-on inclination.
\item These parameters are strikingly consistent with independent constraints from GRAVITY and EHT observations, providing robust evidence that transient hotspots in a nearly face-on accretion disk are the drivers of Sgr A$^*$’s multi-wavelength variability.
\item  GRRT simulations confirm that relativistic corrections and finite hotspot size introduce only minimal deviations from the simple Doppler-beaming model in a low inclination consistent with weak Doppler modulation, reinforcing the robustness of our results.
\end{enumerate}
This study establishes millimeter-wave periodicity as a powerful and transformative probe of the innermost accretion flow around supermassive black holes. By directly connecting ALMA variability studies with infrared and X-ray flares, our findings bridge observational regimes and open a new avenue for exploring the fundamental physics of strong gravity. Future long-term, high-cadence monitoring of Sgr A$^*$ with ALMA, together with advanced relativistic modeling, will be crucial to fully unravel the dynamics of such transient periodicities.

\begin{acknowledgments}
This paper makes use of the following ALMA data: ADS/JAO.ALMA\#2013.1.00111.S, \#2013.1.00764.S, \#2015.1.00311.S, \#2015.1.00859.S, \#2015.A.00021.S, \#2016.1.00870.S, \#2016.A.00037.T, \#2017.1.00503.S, \#2017.1.00820.S, \#2018.1.01124.S, \#2019.1.01559.S, and \#2021.1.00798.S.
ALMA is a partnership of ESO (representing its member states), NSF (USA), NINS (Japan), NRC (Canada), MOST and ASIAA (Taiwan), and KASI (Republic of Korea), in cooperation with the Republic of Chile. The Joint ALMA Observatory is operated by ESO, AUI/NRAO, and NAOJ.
\end{acknowledgments}


\facilities{ALMA}

\software{CASA (version 4.2.2, 4.4.0 4.5.3 4.6.0, 4.7.0, 4.7.2, 5.1.1, 5.4.0, 5.7.2, 6.2.1-7)}

\bibliography{refer_yana}

\begin{thebibliography}{}
\expandafter\ifx\csname natexlab\endcsname\relax\def\natexlab#1{#1}\fi
\providecommand{\url}[1]{\href{#1}{#1}}
\providecommand{\dodoi}[1]{doi:~\href{http://doi.org/#1}{\nolinkurl{#1}}}
\providecommand{\doeprint}[1]{\href{http://ascl.net/#1}{\nolinkurl{http://ascl.net/#1}}}
\providecommand{\doarXiv}[1]{\href{https://arxiv.org/abs/#1}{\nolinkurl{https://arxiv.org/abs/#1}}}

\bibitem[{R. Abuter {et~al.}(2018)Abuter, Amorim, Baub{\"o}ck, Berger, Bonnet, Brandner, Cl{\'e}net, Du~Foresto, De~Zeeuw, Deen, {et~al.}}]{abuter2018detection}
Abuter, R., Amorim, A., Baub{\"o}ck, M., {et~al.} 2018, \bibinfo{title}{Detection of orbital motions near the last stable circular orbit of the massive black hole SgrA,} Astronomy \& Astrophysics, 618, L10

\bibitem[{R. Abuter {et~al.}(2022)Abuter, Aimar, Amorim, Ball, Baub{\"o}ck, Berger, Bonnet, Bourdarot, Brandner, Cardoso, {et~al.}}]{abuter2022mass}
Abuter, R., Aimar, N., Amorim, A., {et~al.} 2022, \bibinfo{title}{Mass distribution in the Galactic Center based on interferometric astrometry of multiple stellar orbits,} Astronomy \& Astrophysics, 657, L12

\bibitem[{R. Abuter {et~al.}(2023)Abuter, Aimar, Seoane, Amorim, Baub{\"o}ck, Berger, Bonnet, Bourdarot, Brandner, Cardoso, {et~al.}}]{abuter2023polarimetry}
Abuter, R., Aimar, N., Seoane, P.~A., {et~al.} 2023, \bibinfo{title}{Polarimetry and astrometry of NIR flares as event horizon scale, dynamical probes for the mass of Sgr A,} Astronomy \& Astrophysics, 677, L10

\bibitem[{K. Akiyama {et~al.}(2022)Akiyama, Alberdi, Alef, Algaba, Anantua, Asada, Azulay, Bach, Baczko, Ball, {et~al.}}]{akiyama2022first}
Akiyama, K., Alberdi, A., Alef, W., {et~al.} 2022, \bibinfo{title}{First Sagittarius A* event horizon telescope results. V. Testing astrophysical models of the galactic center black hole,} The Astrophysical Journal Letters, 930, L16

\bibitem[{B. Aschenbach {et~al.}(2004)Aschenbach, Grosso, Porquet, \& Predehl}]{aschenbach2004x}
Aschenbach, B., Grosso, N., Porquet, D., \& Predehl, P. 2004, \bibinfo{title}{X-ray flares reveal mass and angular momentum of the Galactic Center black hole,} Astronomy \& Astrophysics, 417, 71

\bibitem[{F.~K. Baganoff {et~al.}(2003)Baganoff, Maeda, Morris, Bautz, Brandt, Cui, Doty, Feigelson, Garmire, Pravdo, {et~al.}}]{baganoff2003chandra}
Baganoff, F.~K., Maeda, Y., Morris, M., {et~al.} 2003, \bibinfo{title}{Chandra X-ray spectroscopic imaging of Sagittarius A* and the central parsec of the galaxy,} The Astrophysical Journal, 591, 891

\bibitem[{G. Belanger {et~al.}(2006)Belanger, Terrier, De~Jager, Goldwurm, \& Melia}]{belanger2006periodic}
Belanger, G., Terrier, R., De~Jager, O.~C., Goldwurm, A., \& Melia, F. 2006, \bibinfo{title}{Periodic Modulations in an X-ray Flare from Sagittarius A,} 54, 420

\bibitem[{ {CASA Team et al.}(2022){CASA Team et al.}}]{bean2022casa}
{CASA Team et al.} 2022, \bibinfo{title}{CASA, the Common Astronomy Software Applications for radio astronomy,} Publications of the Astronomical Society of the Pacific, 134, 114501

\bibitem[{A. Eckart {et~al.}(2006)Eckart, Baganoff, Sch{\"o}del, Morris, Genzel, Bower, Marrone, Moran, Viehmann, Bautz, {et~al.}}]{eckart2006flare}
Eckart, A., Baganoff, F., Sch{\"o}del, R., {et~al.} 2006, \bibinfo{title}{The flare activity of Sagittarius A*-New coordinated mm to X-ray observations,} Astronomy \& Astrophysics, 450, 535

\bibitem[{R. Genzel {et~al.}(2003)Genzel, Schödel, Ott, Eckart, Alexander, Lacombe, Rouan, \& Aschenbach}]{genzel_near-infrared_2003}
Genzel, R., Schödel, R., Ott, T., {et~al.} 2003, \bibinfo{title}{Near-infrared flares from accreting gas around the supermassive black hole at the {Galactic} {Centre},} Nature, 425, 934, \dodoi{10.1038/nature02065}

\bibitem[{S. Issaoun {et~al.}(2019)Issaoun, Johnson, Blackburn, Brinkerink, Mo{\'s}cibrodzka, Chael, Goddi, Mart{\'\i}-Vidal, Wagner, Doeleman, {et~al.}}]{issaoun2019size}
Issaoun, S., Johnson, M., Blackburn, L., {et~al.} 2019, \bibinfo{title}{The size, shape, and scattering of Sagittarius A* at 86 GHz: first VLBI with ALMA,} The Astrophysical Journal, 871, 30

\bibitem[{Y. Iwata {et~al.}(2020)Iwata, Oka, Tsuboi, Miyoshi, \& Takekawa}]{iwata2020time}
Iwata, Y., Oka, T., Tsuboi, M., Miyoshi, M., \& Takekawa, S. 2020, \bibinfo{title}{Time variations in the flux density of Sgr A* at 230 GHz Detected with ALMA,} The Astrophysical Journal Letters, 892, L30

\bibitem[{J.~P. McMullin {et~al.}(2007)McMullin, Waters, Schiebel, Young, \& Golap}]{mcmullin2007casa}
McMullin, J.~P., Waters, B., Schiebel, D., Young, W., \& Golap, K. 2007, \bibinfo{title}{CASA architecture and applications,} in Astronomical data analysis software and systems XVI, Vol. 376, 127

\bibitem[{J.~M. Michail {et~al.}(2024)Michail, Yusef-Zadeh, Wardle, Kunneriath, Hora, Bushouse, Fazio, Markoff, \& Smith}]{michail2024multiwavelength}
Michail, J.~M., Yusef-Zadeh, F., Wardle, M., {et~al.} 2024, \bibinfo{title}{Multiwavelength Observations of Sgr A*. II. 2019 July 21 and 26,} The Astrophysical Journal, 971, 52

\bibitem[{M. Miyoshi {et~al.}(2011)Miyoshi, Shen, Oyama, Takahashi, \& Kato}]{miyoshi_oscillation_2011}
Miyoshi, M., Shen, Z.-Q., Oyama, T., Takahashi, R., \& Kato, Y. 2011, \bibinfo{title}{Oscillation {Phenomena} in the {Disk} around the {Massive} {Black} {Hole} {Sagittarius} {A}*,} Publications of the Astronomical Society of Japan, 63, 1093, \dodoi{10.1093/pasj/63.5.1093}

\bibitem[{G.~B. Rybicki \& A.~P. Lightman(1979)Rybicki \& Lightman}]{rybicki2024radiative}
Rybicki, G.~B., \& Lightman, A.~P. 1979, Radiative processes in astrophysics (John Wiley \& Sons)

\bibitem[{M. Subroweit {et~al.}(2017)Subroweit, Garc{\'\i}a-Mar{\'\i}n, Eckart, Borkar, Valencia-S, Witzel, Shahzamanian, \& Straubmeier}]{subroweit2017submillimeter}
Subroweit, M., Garc{\'\i}a-Mar{\'\i}n, M., Eckart, A., {et~al.} 2017, \bibinfo{title}{Submillimeter and radio variability of Sagittarius A*-A statistical analysis,} Astronomy \& Astrophysics, 601, A80

\bibitem[{M.~M. Takahashi {et~al.}(2022)Takahashi, Ohsuga, Takahashi, Ogawa, Umemura, \& Asahina}]{takahashi20223d}
Takahashi, M.~M., Ohsuga, K., Takahashi, R., {et~al.} 2022, \bibinfo{title}{3D photon conserving code for time-dependent general relativistic radiative transfer: CARTOON,} Monthly Notices of the Royal Astronomical Society, 517, 3711

\bibitem[{S.~D. von Fellenberg {et~al.}(2023)von Fellenberg, Witzel, Baub{\"o}ck, Chung, Aimar, Bordoni, Drescher, Eisenhauer, Genzel, Gillessen, {et~al.}}]{von2023general}
von Fellenberg, S.~D., Witzel, G., Baub{\"o}ck, M., {et~al.} 2023, \bibinfo{title}{General relativistic effects and the near-infrared and X-ray variability of Sgr A* I,} Astronomy \& Astrophysics, 669, L17

\bibitem[{S.~D. von Fellenberg {et~al.}(2025)von Fellenberg, Roychowdhury, Michail, Sumners, Sanger-Johnson, Fazio, Haggard, Hora, Philippov, Ripperda, {et~al.}}]{von2025first}
von Fellenberg, S.~D., Roychowdhury, T., Michail, J.~M., {et~al.} 2025, \bibinfo{title}{First mid-infrared detection and modeling of a flare from Sgr A,} The Astrophysical Journal Letters, 979, L20

\bibitem[{G.~C. Weldon {et~al.}(2023)Weldon, Do, Witzel, Ghez, Gautam, Becklin, Morris, Martinez, Sakai, Lu, {et~al.}}]{weldon2023near}
Weldon, G.~C., Do, T., Witzel, G., {et~al.} 2023, \bibinfo{title}{Near-infrared Flux Distribution of Sgr A* from 2005--2022: Evidence for an Enhanced Accretion Episode in 2019,} The Astrophysical Journal Letters, 954, L33

\bibitem[{M. Zechmeister \& M. K{\"u}rster(2009)Zechmeister \& K{\"u}rster}]{zechmeister2009generalised}
Zechmeister, M., \& K{\"u}rster, M. 2009, \bibinfo{title}{The generalised Lomb-Scargle periodogram-a new formalism for the floating-mean and Keplerian periodograms,} Astronomy \& Astrophysics, 496, 577

\end{thebibliography}
\bibliographystyle{aasjournalv7}

\end{document}